\documentclass[%
 reprint,
 amsmath,amssymb,
 aps,
]{revtex4-1}

\usepackage{graphicx}% Include figure files
\usepackage{dcolumn}% Align table columns on decimal point
\usepackage{bm}% bold math

\begin{document}

\preprint{APS/123-QED}

\title{TiDeH: Time-Dependent Hawkes Process for Predicting Retweet Dynamics}  % Force line breaks with \\
%  Manuscript Title:\\with Forced Linebreak}

\author{Ryota Kobayashi}
 \email{r-koba@nii.ac.jp}
 \affiliation{Principles of Informatics Research Division, National Institute of Informatics, 2-1-2 Hitotsubashi, Chiyoda-ku, Tokyo, Japan}
\affiliation{Department of Informatics, Graduate University for Advanced Studies (Sokendai), 2-1-2 Hitotsubashi, Chiyoda-ku, Tokyo, Japan}

\author{Renaud Lambiotte}
\affiliation{Department of Mathematics and naXys, University of Namur, 8 Rempart de la Vierge, Namur B-5000, Belgium}

\date{\today}% It is always \today, today,

\begin{abstract}
Online social networking services allow their users to post content in the form of text, images or videos. The main mechanism driving content diffusion is the possibility for users to re-share the content posted by their social connections, which may then cascade across the system. A fundamental problem when studying information cascades is the possibility to develop sound mathematical models, whose parameters can be calibrated on empirical data, in order to predict the future course of a cascade after a window of observation. 
In this paper, we focus on Twitter and, in particular, on the temporal patterns of retweet activity for an original tweet. We model the system by Time-Dependent Hawkes process (TiDeH), which properly takes into account the circadian nature of the users and the aging of information. The input of the prediction model are observed retweet times and structural information about the underlying social network. We develop a procedure for parameter optimization and for predicting the future profiles of retweet activity at different time resolutions. We validate our methodology on a large corpus of Twitter data and demonstrate its systematic improvement over existing approaches in all the time regimes.
\end{abstract}

\pacs{Valid PACS appear here}% PACS, the Physics and Astronomy
\maketitle

%\tableofcontents

\section{Introduction}

In recent years, online social networking sites (OSNs) have become an increasingly central medium for information diffusion. In OSNs, users can generate their own content, but also discover information generated by their social contacts and re-share it to their own contacts. Importantly, an information can be re-shared multiple times, and the resulting multiplicative mechanism may lead to cascades over a large number of people, possibly even reaching regions of the social graph distant from the original post \cite{Kwak2010}. Such cascades have been identified in a variety of OSNs, including Facebook and Twitter \cite{Dow2013,Kumar2010}.

A growing body of research has improved our understanding of information cascades, from the design of accurate theoretical models for diffusion \cite{Easley2010,Goetz2009,Centola2010}, to the empirical study of the structural properties of cascades \cite{Adar2005,Gruhl2004,Salah-Brahim2012} and of their interplay with the structure of the underlying topology \cite{Weng2014}. From a practical point of view, a crucial question is to predict the future evolution of an information cascade, based on observations made during its early stage. The most simple way to formulate this problem is to predict the final size of the information cascade, that is the total number of direct and indirect re-shares received by a given post. This prediction problem has important applications for the good-functioning of OSNs, for instance to rank content and improve the presentation of information to often overflowed users, and for media campaign management. In a machine learning framework, this problem can be solved as a classification task, where an exhaustive set of features, including semantic, structural and temporal information, are fed into standard classification methods \cite{petrovic2011rt,Hong2011,Bao2013,cheng2014}. An alternative approach consists in building realistic, yet simple and principled, models of information diffusion, and fitting their parameters on empirical data \cite{zaman2014,gao2015,zhao2015seismic}. This modeling approach has the advantage of improving our understanding of the mechanisms driving diffusion, and of testing the predictive power of information diffusion models.

\textbf{Present work:} In this paper, we extend the classical problem of cascade size prediction and aim at predicting how the cascade size evolves in time.  In practice, we focus on Twitter and on the number of retweets of an original tweet, but our method is general and can be applied to any type of OSN. Our problem is the following: given a time series of retweets during a window of observation, being able to predict the time evolution of the frequency at which retweets will appear in the future, at different temporal resolutions. We are thus interested in  predicting not only a  number, the final size of the cascade, but a curve, how popularity will evolve in time, after a window of observation. To do so, we adopt a modeling perspective and see the time series as a Time-Dependent Hawkes process, TiDeH, which generalizes a classical model for self-exciting point processes. Hawkes process differ from memoryless Poisson processes as the future rate of activity is boosted by the occurrence of previous events \cite{hawkes1971}. They can themselves be seen as generalizations of epidemiological models and branching processes \cite{Newman2010}, where an additional ingredient is incorporated, a memory kernel determining the time between a cause, e.g. a tweet, and its effect, a retweet. Hawkes processes have been adopted in a wide range of applications, including information diffusion in OSNs \cite{zhao2015seismic}, where their multiplicative nature naturally translates the fact that a new re-share exposes new followers and may thus provoke new re-shares in the future. Here, we additionally make the Hawkes process time-dependent by allowing the model parameter to vary daily. 

\begin{figure}[!h]
  \centering
  \includegraphics[width=7cm]{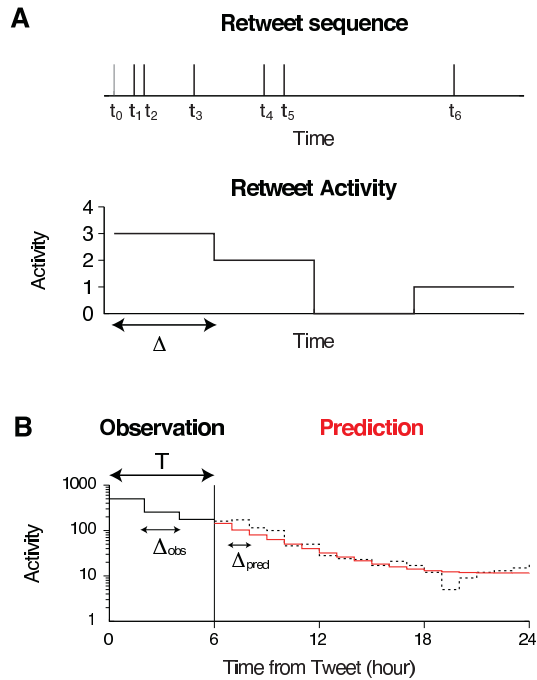}
  \caption{Predicting retweet activity. A. Top: Retweet sequence. Black (Grey) bars represent the retweet (original tweet) times. Bottom: Retweet activity. Retweet activity is defined by the number of retweet in a time window.  The duration of the time bin $\Delta$ determines the temporal resolution of the analysis.
B. Prediction problem. We aim to predict the future retweet activity from observed retweet times and number of followers up to time $t_0 + T$. In general, two different types of binning, $\Delta_{\rm obs}$ and $\Delta_{\rm pred}$, can be used for the estimation of model parameters and for the prediction of future events.}
% We aim to predict the future retweet activity from observed retweet times and  number of followers up to time $t_0+T$.}
  \label{fig:Schem_Predict}
\end{figure}

Our methodology is set up as follows: for a tweet of interest, we observe its retweet sequence  $\{t_i, d_i \}$ up to time $t_0+ T$, where $t_i$ is the $i$-th retweeted time, $d_i$ is the number of followers of the $i$-th retweeting person, $t_0$ is the posted time of the original tweet,  $d_0$ is the number of followers of the tweeting person, and $T$ is the duration of the observation. 
We first fit the parameters of TiDeH based on time series of retweets and information on the number of followers.  We then predict retweet activity, defined as the number of retweets in the $k$-th bin $ t \in [(k-1) \Delta_{\rm pred}, k \Delta_{\rm pred} \}$, where $\Delta_{\rm pred}$ is the bin width that represents the time resolution of the prediction (Fig.~\ref{fig:Schem_Predict}). Let us note here that our prediction task generalizes the task of predicting the total number of retweets, as the latter is recovered either for large values of $\Delta_{\rm pred}$ or by summing all of the future values of retweet activity.
The prediction of future events is performed by solving numerically a self-consistent integral equation of the model during the prediction period. As we will see, TiDeH presents a series of advantages over existing approaches, as it significantly improves the accuracy of predictions and that it provides a systematic, mathematically sound, framework to predict temporal variations of re-shares in OSNs.

 \vspace{0.5cm}

The rest of the paper is organized as follows: Section 2 surveys
the related work. In Section 3, we describe the data-set, describe TiDeH, and provide evidence for the circadian dependence of the model parameter. We then devise an optimization method for parameter estimation and test it on artificial data. In section 4, we present our procedure to predict the future profiles of retweet activity and compare the accuracy of different versions of TiDeH. In section 5, we thoroughly evaluate our method on empirical data and compare its performance
with state-of-the-art approaches. In Section 6, we conclude and discuss future research directions.

 \vspace{3cm}
%%  
%%   Section 2
%%
\section{Related work}

The study of information cascades in OSNs is an active field of research \cite{Rogers2010}. Many papers have analyzed and described the temporal and structural properties of empirical information cascades \cite{Dow2013,Kumar2010}. In parallel, theoretical works have considered the design of theoretical models of cascade dynamics in networks \cite{Easley2010}. Our work is at the interface between these approaches, as the prediction of the future course of a cascade is performed through a  properly calibrated information diffusion model. The problem of cascade prediction is generally defined to estimate the final size of a cascade, or equivalently the total popularity of an original post. Broadly speaking, two types of methods have been developed to solve this problem. 
On the one hand, machine learning methods consist in collecting an exhaustive list of potentially relevant features for each cascade, including semantic content, meta-information, structural and temporal features. Learning or statistical methods are then applied in order to classify the cascades and predict their future size. Following the seminal  observation  that popularity on early days and later ones  are high log-linear correlated \cite{szabo2010}, more recent works focusing on Twitter include \cite{petrovic2011rt}, where learning techniques are shown to achieve similar performance to humans, but also  \cite{yang2010} showing that  the number of \@username mentions helps predicting the speed and shape of retweet dynamics. In the case of re-shares on Facebook, let us also mention \cite{cheng2014} where the authors  observe that temporal and structural features of cascades are key predictors for their growth. Known drawbacks of this family of methods include a high sensitivity to the quality of the features, the requirement of an extensive training,  and thus a limited applicability in real-time online settings \cite{Bandari2012}.

A second type of predictive methods aims at calibrating models of information diffusion on time series of events during a window of observation, possibly by incorporating additional social network information. Two important ingredients of the models are the instantaneous character of events in OSNs,  which are thus so-called ``point processes" in the mathematical literature, and the multiplicative nature of the diffusion, as new events tend to trigger new ones. For these reasons, several works have developed models based on self-exciting point processes and, in particular, Hawkes processes. A major distinction between our model and existing ones \cite{zaman2014,Yang2013} is its time-dependence as we take into account circadian rhythms of online popularity and aging of information. Importantly, the infectiousness of the original tweet  naturally depends on its posting time, in agreement with observations that it is an intuitive predictor for popularity \cite{petrovic2011rt}, and this effect is also present for later retweets.
Our model can be seen as a time-dependent extension of SEISMIC  \cite{zhao2015seismic}, as our model also incorporates a partial information of the network structure, but with two additional  differences. First, our goal is to predict the time evolution of the number of retweets in the future, and not simply the total number of retweets. As we show below, the incorporation of circadian patterns is  particularly important to improve accuracy in this context.  Second, we develop a framework for predicting future activity that is  mathematically consistent with the modeling. Our model is also related to the time-dependent Poisson process model \cite{gao2015}, which we compare to TiDeH below, and to SpikeM \cite{matsubara2012}, which incorporates daily cycles and a finite population ensuring an asymptotic decay of the propagation, but differs from our approach by its deterministic and descriptive character. 
These two models also have the drawback of neglecting the effect of social network topology on information cascade, despite its important impact in spreading processes. 

Beyond these works on information diffusion, it is important to emphasize here that Hawkes processes have been applied in a variety of settings in order to describe and to predict univariate or multivariate data. Originally defined to describe earthquake dynamics \cite{hawkes1971}, where a power-law memory kernel was first introduced \cite{ogata1988}, it was for instance applied to predict where and when aftershocks would occur \cite{helmstetter2003}. In finance, and in particular high frequency finance \cite{bacry2015}, estimations of the model parameters allow to quantify if price changes are dominated by endogenous feedback processes, as opposed to exogenous news \cite{filimonov2012}. Similar applications have also been developed to model popularity of online content, in particular in Youtube \cite{crane2008}, by estimating the different types of response after endogenous and exogenous bursts of activity. In social dynamics, \cite{masuda2013} showed that the model can help reproduce empirical features observed in conversation event sequences, and \cite{Mohler2011} applied it in order to predict criminal events. In scientometrics, citation dynamics have been also modeled by modified Hawkes models \cite{michael2012}, and future citations of a given paper predicted by reinforced Poisson process \cite{Shen2014}. Finally, let us also note that Hawkes processes have also triggered  theoretical research associated to the non-Markovian nature of their dynamics, and its impact on spreading times \cite{Delvenne2015}. 
%  \input{sec2-relatedwork.tex}

%%  
%%   Section 4
%%       \input{sec4-modeling.tex}
\section{Modeling retweet activity via time-dependent Hawkes Process}

\subsection{Data sets}
We analyzed 166,076 tweets on Twitter from October 7 to November 7, 2011, which was used in a previous study~\cite{zhao2015seismic} and available in \url{http://snap.stanford.edu/seismic/}. 
For each tweet, the dataset includes tweet ID, posting time, time of retweets, and the number of followers of  users for the original tweet and later retweets.  The retweet times are recorded up to 7 days (168 hours) 
% \textcolor{blue}{
from the original post for each tweet. 
Note that the data contains some minimal information  about the network structure  (the number of followers), as it is easily available through the Twitter API, but the presence of connections between users in the Twitter network  is not known. 
We focus on a subset of popular tweets (738 tweets) that have at least 2,000 retweets in order to calibrate our model and to evaluate the performance of our predictions. 
\subsection{TiDeH: Time-dependent Hawkes process}
\begin{figure}[h!]
  \centering
  \includegraphics[width=7cm]{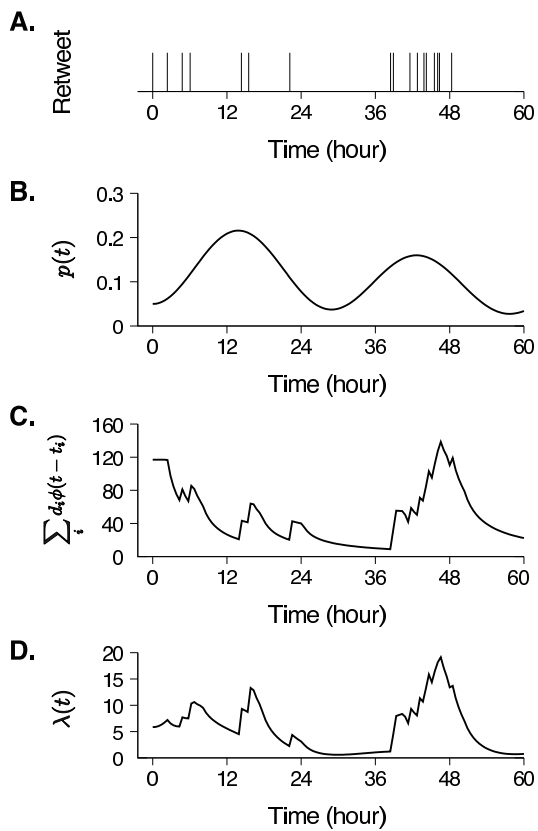}
  \caption{Time-dependent Hawkes process. A. Simulated retweet activity. The original tweet (time 0) and retweet times are represented as a bar. B. Infectious rate $p(t)$ (1st term of Eq.~\ref{eq:Self-Exc}). C. Memory Effect $\sum_i d_i \phi(t-t_i)$ (2nd term of Eq.~\ref{eq:Self-Exc}). D. Instantaneous probability of a retweet $\lambda(t)$, as obtained by the product of B and C.}
  \label{fig:TD_Hawkes}
\end{figure}

We develop a Time-Dependent Hawkes process (TiDeH) for predicting retweet activity, and extending the classical stationary Hawkes process~\cite{hawkes1971,zhao2015seismic} (Fig.~\ref{fig:TD_Hawkes}). %  We consider a tweet posted at $t_0=0$.
The probability for getting a retweet in a small time interval $[t, t+ \Delta t]$ is described as 
\begin{equation}
        \textrm{Prob} \left( \textrm{Getting a retweet in }\ [t, t+ \Delta t] \right)= \lambda(t) \Delta t,  \label{eq:Poisson}
\end{equation} 
where the time-dependent rate depends on previous events as
\begin{equation}
	\lambda(t) = p(t) \sum_{i: t_i<t} d_i \phi(t-t_i), \label{eq:Self-Exc}
\end{equation} 
and where $p(t)$ is the infectious rate, $t_i$ is the time of $i$-th retweet. Following \cite{zhao2015seismic}, we also incorporate  the number $d_i$ of followers of the $i$-th retweeting person. By doing so, the model essentially generates a branching process for the diffusion, and gives more importance to highly connected nodes. This step is akin to tree-like and heterogeneous mean-field approximations popular to simplify the theoretical study of epidemic spreading on networks \cite{Newman2010}. 
The memory kernel $\phi(s)$ is a probability distribution for the reaction time of a follower, that is the time interval between a tweet by the followee and its retweet by the follower. This distribution has been shown to be heavily tailed in a variety of social networks \cite{vazquez2006,crane2008}, and it is fitted to the empirical data by the function
\begin{equation}
  \phi(s) =  \begin{cases}
    0    & (s<0) \\
    c_0  & (0 \leq s \leq s_0) \\
    c_0 (s/s_0)^{-(1+\theta)}  & (\textrm{Otherwise}) \\
  \end{cases}, 
\end{equation} 
The parameters were set to $c_0= 6.49 \times 10^{-4}$ (/seconds), $s_0= 300$ seconds, and $\theta= 0.242$ \cite{zhao2015seismic}. % As compared to a classical Hawkes model, the main contribution  and, as we will see, source of computational complexity of our work is the time-dependent infectious rate $p(t)$. 
As we show in the following section, $p(t)$ is observed to decrease to zero for sufficiently long times, which ensures that the predicted number of retweets does not diverge. 

\vspace{0.5cm}

\newpage
\subsection{Modeling the infectious rate of a tweet}   \label{sec:fit_pt}
\subsubsection{Estimating the instantaneous infectious rate.}
\begin{figure}[h!]
  \centering
  \includegraphics[width=7cm]{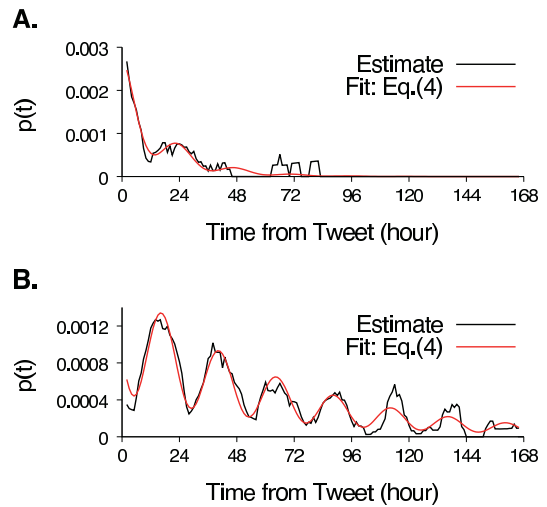}
  \caption{Estimated infectious rate from two retweet sequences. Two types of  dynamics are observed, i.e., a decay (A) and a decay with circadian oscillations (B). Black lines are the  rates estimated by the moving time window and red lines indicate the fit by the proposed model.}
  \label{fig:Ex_Est_pt}
\end{figure}

The infectious rate $p(t)$ is estimated by using moving time windows. Assuming that the infectious rate is constant in a small time window $t \in [t_{\textrm{st} }, t_{\textrm{en} }]$, $p(t)$ is calculated by the maximum likelihood method, 
\begin{equation}
%     \hat{p}_t= \frac{\delta R}{ \sum_i n_i \{ \Phi(t_{\textrm{en} }-t_i)- \Phi(t_{\textrm{st} }-t_i) \} },   \label{eq:Estp0_ML}
     \hat{p}_t= \frac{\delta R}{ \sum_i n_i \{ \Phi(t_{\textrm{en} }-t_i)- \Phi(t_i- t_{\textrm{st} }) \} },   \label{eq:Estp0_ML}
\end{equation} 
where $\delta R$ is the number of retweets in the time window and $\Phi(t)$ is the integral of the memory kernel, $\Phi(t)= \int_0^t \phi(s) ds $. 
Note that alternative methods could be applied, without the need for moving time windows, for instance by using the empirical Bayes method~\cite{Koyama2005}. However, our choice is motivated by its simplicity and  the window size ($\Delta_{\rm obs}= t_{\textrm{en} }- t_{\textrm{st} }$) is set to 4 hours. 
Examples of the estimated infectious rate from a retweet sequence are shown in Figure~\ref{fig:Ex_Est_pt}. 

\subsubsection{Modeling the infectious rate of retweets.}
Infectious rates $p(t)$ from a retweet sequence (Fig.~\ref{fig:Ex_Est_pt}) clearly show two properties, a circadian cycles and a slow decay. The decay is expected due to the inevitable aging of information, whose life-cycle is known to be short in microblogging systems, but also to the decreasing number of potentially interested followers, as the cascade progresses. The oscillations are expected for cascades that remain geographically localized, within a limited number of time zones, such that daily cycles of human activity naturally translate into cycles of retweet activity.
Based on this observation, we propose a minimal model for the time dependence of the infectious rate
\begin{equation} 
    p(t)= p_0 \left\{ 1 - r_0 \sin \left( \frac{2\pi}{T_m} (t+ \phi_0) \right) \right\} e^{- (t-t_0)/\tau_m  } \label{eq:TiDe_pt}
\end{equation} 
where, as before, $t_0$ is the time of the original tweet.  The period of  oscillation is set to $T_m= 1$ day.  The parameters, $p_0, r_0, \phi_0, \tau_m$ correspond to the intensity, the relative amplitude of the oscillation, its phase, and  the characteristic time of popularity decay respectively. These four parameters are fitted by minimizing the least square error 
\begin{equation}    
    E_p= \sum_{k=1}^M \{ \hat{p}_k - p( (k+0.5) \Delta_{\rm obs}) \}^2, 
\end{equation}    
where $M= T/\Delta_{\rm obs}$ is the number of bins, and $\hat{p}_k$ is the estimate of the infectious rate in a time bin $t \in [k \Delta_{\rm obs}, (k+1) \Delta_{\rm obs} ]$. The Levenberg$-$Marquardt algorithm is then used to minimize the error, and the parameter range of $r$ and $\tau_m$ is restricted, i.e., $-1<r<1$ and $0.5 < \tau_m < 20$ days. 
\subsubsection{Validation of the fitting procedure on synthetic data.} 
\begin{figure}[t]
  \centering
  \includegraphics[width=7cm]{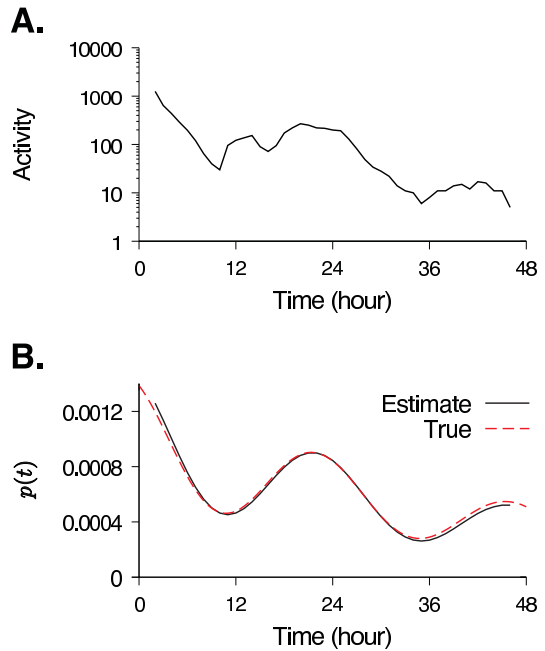}  
  \caption{Estimating the infectious rate $p(t)$ from synthetic data. A. Activity of simulated data. The number of events in two hours window is counted. B. Fitted infectious rate. The rate of the model with optimized parameters (black) is compared to the true rate (red dashed).}
  \label{fig:Ex_SimData}
\end{figure}

We validate the fitting procedure for $(p_0, r_0, \phi_0, \tau_m)$ by analyzing synthetic data generated by TiDeH (\ref{eq:Poisson},\ref{eq:Self-Exc}) with the time-dependent infectious rate (\ref{eq:TiDe_pt}). The number of followers $d_i$ was obtained from a retweet sequence of the empirical data. 
Figure \ref{fig:Ex_SimData} shows that the fitting procedure, when applied to one retweet sequence, can reconstruct the unobservable infectious rate from the simulated sequence.
We evaluate the accuracy of the parameter estimation by comparing its estimates with  the ``ground-truth'' values used to generate the synthetic data. 
Table 1 summarizes the mean and standard deviation of the estimates for 100 trials. The relative errors are 0.0 \%, 1.9 \%, 9.6\%, and 1.0 \%, for $p_0$, $r_0$, $\phi_0$, and $\tau_m$, respectively, suggesting that the fitting procedure accurately reconstructs the parameters for sufficiently long observation period, here set to $T=$ 2 days. 

%  Table 1:  Estimated parameters (Simulated data)
\begin{table}[htb]
  \caption{Parameter estimation by the least square method from simulated data (observation time $T$: 2 days ). }
  \vspace{3mm}
  \begin{tabular}{|l||r|r|} \hline
    Parameter  &  Estimate  &  True   \\ \hline 
    $p_0$      &  ${\bf 0.001 \pm 0.00009}$  &  $0.001$    \\
    $r_0$      &  ${\bf 0.416 \pm 0.069}$  &   $0.424$    \\
    $\phi_0$   &  ${\bf 0.113 \pm 0.030}$  &   $0.125$    \\ \hline
    $\tau_m$   &  ${\bf 2.02 \pm 0.66}$    &   $2.00$      \\  \hline
  \end{tabular}
\end{table}

%  Fig.4: Application to Simulated data (Example)
\vspace{3cm}

\begin{figure}[h!]
  \centering
  \includegraphics[width=8cm]{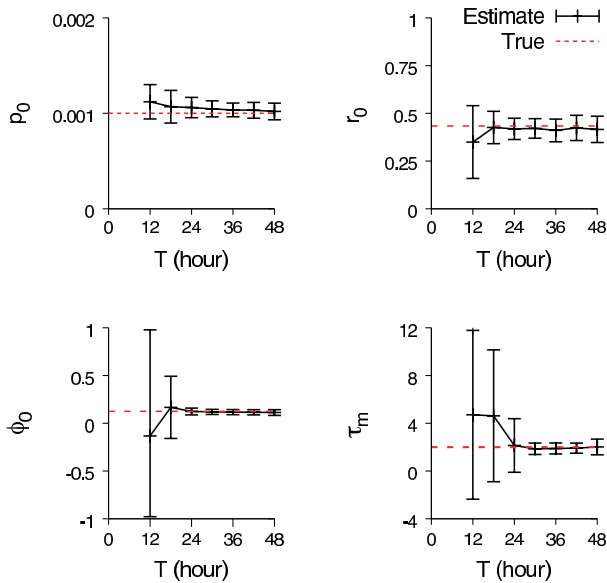}
  \caption{Dependence of the accuracy of parameter estimation on the observation time $T$. The mean and standard deviation of the estimates are calculated from 100 synthetic data.}
  \label{fig:Accuracy_Full}
\end{figure}

\begin{figure}[h!]
  \centering
  \includegraphics[width=4cm]{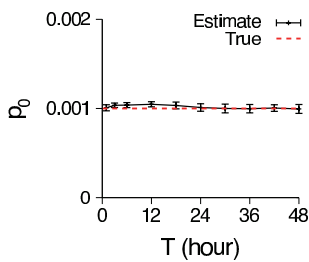}
  \caption{Dependence of the accuracy of parameter estimation on the observation time $T$: case when only $p_0$ is unknown.}
  \label{fig:Accuracy_Single}
\end{figure}

%  Short observation time
As a next step, we examine the dependence of the estimation accuracy on the duration of the observation period.   Fig.~\ref{fig:Accuracy_Full} shows that  accuracy deteriorates for short durations. 
In particular, we cannot obtain  reliable estimates for the phase $\phi_0$ and the time constant $\tau_m$ if the observation time is shorter than 24 hours. 
% Single parameter case 
A possible reason for this lack of accuracy is that 
%\textcolor{blue}{
there are too many parameters to be estimated from limited data. 
 To test this hypothesis, we consider the situation when only the amplitude parameter $p_0$ is to be fitted, while  the other parameters ($r_0, \phi_0, \tau_m$) are known. In that case, $p_0$ can be accurately estimated, even from a very short observation window, $T= 1$ hour (Fig.~\ref{fig:Accuracy_Single}).

%  Fig.5: Accuracy of the estimate (1)

%  Fig.6: Accuracy of the estimate (2)
\vspace{5cm}

%%    
%%    Section  5: 
%%   	\input{sec5-predHawkes.tex}
\section{Predicting future retweet activity via TiDeH} 
We develop a procedure to predict the future retweet activity of an original tweet based on TiDeH. It consists in two steps.
First, the infectious rate $p(t)$ is calibrated. 
Second, the future retweet rate $\lambda(t)$ is calculated based on the infectious rate and the observed retweet sequence $\{ t_i, d_i \}$ $(t_i<T)$, and the future retweet activity is estimated.
\subsection{Step 1: Fitting the infectious rate $p(t)$}     \label{sec:pred_fit_pt}
We consider three ways to identify the infectious rate $p(t)$ from a retweet sequence. 
%  1st approach: Constant rate $p(t)= p_0$
In a first approach, we assume that the infectious rate is constant $p(t)= p_0$ and this single parameter is estimated from the observed retweet sequence by the maximum likelihood method (\ref{eq:Estp0_ML}). 
%  2nd approach: Time-dependent rate $p(t)= p_0 (1- r \sin( 2\pi (t+\phi_0) ) ) e^{-(t-t_0)/\tau_m }$, Estimate all the parameters from the target retweet sequence.
In a second approach, we consider the model (\ref{eq:TiDe_pt}) for the time-varying infectious rate, and all the parameters $(p_0, r_0, \phi_0, \tau_m)$ are estimated from the observed retweet sequence by using the fitting procedure developed in section~\ref{sec:fit_pt}. 
%  3rd approach: Time-dependent rate $p(t)= p_0 (1- r \sin( 2\pi (t+\phi_0) ) ) e^{-(t-t_0)/\tau_m }$, Estimate only the amplitude parameters from the target 
In a third approach, we again adopt (\ref{eq:TiDe_pt})  for the time-varying infectious rate, but we now optimize the shape parameters $(r_0, \phi_0, \tau_m)$ by minimizing the prediction error (Sec.~\ref{sec:comp_Hawkes}) on a training data set. We used the simplex downhill method~\cite{nelder1965} for the minimization. Then, the intensity $p_0$ was estimated from the retweet sequence of interest by using the fitting procedure developed in section~\ref{sec:fit_pt}. This method with training is motivated by our observation that the prediction for $p_0$ is accurate when the other 3 parameters are fixed, even for short observation windows, and its  performance was evaluated by using a 5-fold cross validation. From now on, we call the models associated to the three different fitting procedures  standard Hawkes process, TiDeH without training  and TiDeH with training respectively.
\subsection{Step 2: Evaluating the future retweet activity} 
The retweet activity $A_k$ is defined as the number of retweets in the $k$-th bin and it is determined from the retweet rate $\lambda(t)$ by (\ref{eq:Poisson}). 
To calculate the retweet rate $\lambda(t)$, we need to know all the previous retweet times $t_i$ up to time $t$. Unfortunately, we can observe the retweet times only up to time $T$. 
To incorporate the impact of unobserved retweets after time $T$, we consider the expectation of the retweet rate given the $R(T)$ retweet times up to time $T$, 
\begin{equation}
  	\hat{\lambda}(t) = E[ \lambda(t) | t_1, t_2, \cdots , t_{R(T)} ].
\end{equation} 
Taking the conditional expectation on Eq.(\ref{eq:Self-Exc}), a self-consistent equation can be derived as 
\begin{equation}
       	\hat{\lambda}(t) =  f(t)+ d_p p(t) \int_T^t \hat{\lambda}(t) \phi(t-s) ds,    \label{Eq.Volttera-eq}
\end{equation}
where we assumed that the random variables for $d_i$ and $t_i$ are independent, and 
\begin{equation}
  f(t)= p(t) \sum_{i: t_i<T} d_i  \phi(t-t_i).
\end{equation}
$d_p$ is the conditional expectation of $d_i$ for $i>R(T)$,
\begin{equation}
  d_p= E[ d_i | t_1, t_2, \cdots , t_{R(T)} ],
\end{equation} 
and it is estimated by the mean number of  followers during the observation window. 
The first term of (\ref{Eq.Volttera-eq}) describes the contribution of the observed retweets and the second term describes that of the self-excitation induced during the prediction period. Eq. (\ref{Eq.Volttera-eq}) is known as a Volttera integral equation, and it can be numerically solved by evaluating the integral by the trapezoidal method~\cite{press1996numerical}. Here, we set the time step to 0.1 hour.

An alternative approach to  evaluate the future retweet rate $\lambda(t)$ consists in performing  Monte Carlo simulations of  TiDeH for a number of realizations, and in calculating the average value of $\lambda(t)$.  We did not adopt this approach, because it requires a high computational cost to generate sufficiently large samples of the stochastic process. When comparing the two approaches, we have found that at least 10,000 realizations of the Monte Carlo simulations are required to produce reasonable estimates for the retweet rate $\lambda(t)$ (Data not shown).
\subsection{Effect of the infectious rate models on prediction performance}  \label{sec:comp_Hawkes}
Let us now examine how the choice of fitting procedure for the infectious rate, described in subsection \ref{sec:pred_fit_pt}, impacts the prediction performance. 
The quality of the prediction is evaluated by the mean and by the median of the absolute error. 
The absolute error per hour is defined as 
\begin{equation}
   \epsilon_A= \frac{1}{T_{\rm max}-T} \sum_k |\hat{A}_k- A_k|, 
\end{equation}
where $\hat{A}_k$ and $A_k$ are the predicted and actual value of the retweet activity in the $k$-th bin, 
% \textcolor{blue}{ 
and $T_{\rm max}= 168$ hour is the end time of prediction period. 

\begin{figure}[h!]
  \centering
    \includegraphics[width=8cm]{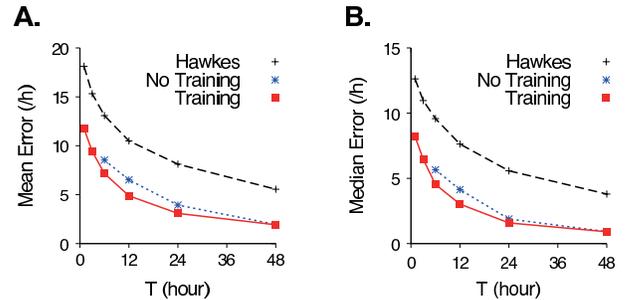}
  \caption{Dependence of the error (A: Mean, B: Median) on observation time $T$, for the three proposed models: standard Hawkes process, TiDeH without traning and TiDeH with traning. The window size $\Delta_{\rm pred}$ is set to 4 hours.} 
  \label{fig:Comp_Hawkes_T}
\end{figure}

\begin{figure}[h!]
  \centering
    \includegraphics[width=8cm]{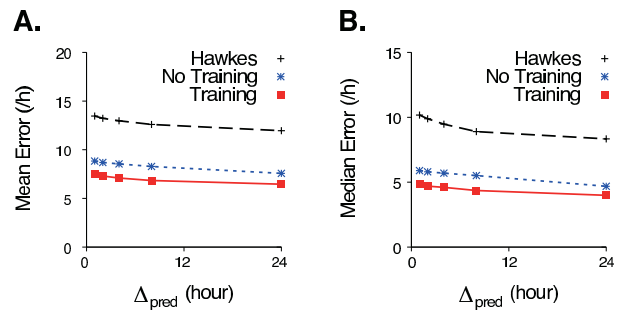}
  \caption{Dependence of the error (A: Mean, B: Median) on time resolution $\Delta_{\rm pred}$. Abbreviations of the prediction methods are the same as in Fig.~\ref{fig:Comp_Hawkes_T}. The observation time $T$ is fixed to 6 hours. }
  \label{fig:Comp_Hawkes_DT}
\end{figure}

%  Prediction performance 1
We first consider the effect of the observation time on prediction performance (Fig.~\ref{fig:Comp_Hawkes_T}). 
%  1. TiDeH is better than Hawkes
TiDeH clearly outperforms the standard Hawkes model for all values of $T$.  For example, the median error of TiDeH with training is 8.2  for $T=1$ hour and 1.6 for $T=1$ day, to be compared with 12.6 and 5.6 for the standard Hawkes process respectively. As expected, longer observation windows improve the accuracy of the predictions.
%  2. Training improves!
We also observe that training can improve the prediction performance for short observation windows ($T< 24$ hours), and that the model with training provides accurate predictions, even for very short observation windows, such as $T= 1$ hour. The model without training is accurate for sufficiently large values of $T$, but it cannot be applied for short observations because the quality of parameter fitting deteriorates, as we showed in Sec.~\ref{sec:fit_pt}.  
%  Prediction performance 2
Finally, we consider the effect of the time resolution $\Delta_{\rm pred}$, that is the granularity of the time dependence, on prediction performance (Fig.~\ref{fig:Comp_Hawkes_DT}). TiDeH again performs significantly better than the standard Hawkes model, and its error is roughly independent of the time resolution.  Overall, these results show that TiDeH with training is the best predictive method among the three methods, and it is thus selected for comparison  to state-of-the-art methods  in the next section.

\vspace{1cm}
\subsection{Summary of TideH with training}
Our selected procedure to predict  future retweet activity  is summarized in table~\ref{table:pred_method}. Given a desired value of temporal resolution $\Delta_{\rm pred}$, we proceed as follows:
First, we identify the infectious rate of a tweet $p(t)$ by fitting the proposed oscillatory model. We recommend to optimize the shape parameters ($r, \phi_0, \tau_m$) using a training data set, and then to estimate the intensity $p_0$ for the target retweet sequence. 
Second, we calculate the mean number of followers in the target sequence. 
Third, the time course of the future retweet activity $\hat{\lambda}(t)$ is evaluated by solving numerically the self-consistent equation for TiDeH (\ref{Eq.Volttera-eq}). 
Finally, the retweet activity in a bin $A_k$ is calculated from the estimated retweet rate, 
$A_k= \int_{T+\Delta_{\rm pred} (k-1)}^{T+\Delta_{\rm pred} k} \hat{\lambda}(s) ds$. 
The computational cost after parameter optimization is $O(R(T) T_{\rm pred} ) + O(T^2_{\rm pred})$ where $R(T)$ is, as before, the number of observed events, and $T_{\rm pred}$ is the duration of the prediction period.

\begin{table}[htb]
  \caption{\textbf{Selected method for the prediction of future retweet activity (TiDeH with training)} }
  \vspace{3mm}
  \begin{tabular}{l}    
    \textbf{1. Identify the infectious rate $p(t)$.} \\
       \quad a) Optimize the shape parameters $(r_0, \phi_0, \tau_m)$ by \\ 
       \quad \quad \quad   minimizing the error for the training data. \\
       \quad  b) Fit the amplitude $p_0$ from the retweet sequence. \\ \\     
    \textbf{2. Calculate the average number of followers.} \\ \\ 
    \textbf{3. Evaluate future retweet rate $\hat{\lambda}(t)$ by solving} \\ 
    \quad \textbf{the integral equation (\ref{Eq.Volttera-eq}).}  \\ \\ 
    \textbf{4. Evaluate the mean number of retweet $A_k$.} \\ 
    \quad \quad \quad  $A_k= \int_{T+(k-1) \Delta_{\rm pred} }^{T+k \Delta_{\rm pred}} \hat{\lambda}(s) ds.$
  \end{tabular}
  \label{table:pred_method}
\end{table}
%%
%%    
%%    Section  6
%%   	\input{sec6-comparison.tex}
\section{Comparison of prediction performance with previous methods}

\subsection{Baseline methods for comparison}
In this section, we describe four  methods used  as a baseline to estimate the predictive performance of our method. 
It should be noted that a direct comparison can not be performed because previous methods were originally designed for  different prediction tasks: our work predicts the time evolution of retweet activity, whereas  previous works \cite{szabo2010,zhao2015seismic} primarily focused on predicting the final number of retweets. 
For this reason, we have modified  three  of the existing methods so that they now predict the cumulative number of retweet up to time $t$, $R(t)$. The number of retweets in the $k$-th bin $(t \in [T+(k-1) \Delta_{\rm pred}, T+k \Delta_{\rm pred}])$ can then be calculated from the cumulative number of retweets by
\[    A_k= R(T+k \Delta_{\rm pred})- R \left( T+(k-1) \Delta_{\rm pred} \right). \] 
The fourth method is only used to evaluate the accuracy of TiDeH to predict the final number of retweets in the next section.
\subsubsection{Linear regression (LR)~\cite{szabo2010}.}
The first method is  a linear regression of the logarithm of the popularity $R(t)$ performed on a training set of $n_{\rm tr}$ tweet sequences
\[
        \log R(t)= \alpha_t + \log R(T) + \sigma_t \xi_t.
\] 
 $\alpha_t$ is  obtained by minimizing the squared error 
\[	
        E_t(\alpha_t)=  \sum_{k=1}^{n_{\rm tr} } \left\{ \log R_k(t) - \alpha_t- \log R_k(T) \right\}^2,
\] 
and $R_k(t)$ is the cumulative number of retweets for the $k$-th tweet in the training data and $\xi_t$ is a gaussian random variable with zero mean and unit variance. 
The variance $\sigma_t$ is determined by the maximum likelihood estimator $\hat{\sigma}_t^2= E_t(\hat{\alpha}_t)/ n_{\rm tr}$, where $\hat{\alpha}_t$ and $\hat{\sigma}_t^2$ are the fitted values of $\alpha_t$ and $\sigma_t^2$ respectively. 
The cumulative number of retweets $R(t)$ is predicted by the unbiased estimator
\[	
        \hat{R}(t)= R(T) \exp( \hat{\alpha}_t + \hat{\sigma}^2_t/2). 
\] 
\subsubsection{Linear regression with degree (LR-N)~\cite{zhao2015seismic}.}
The second method is an extension of the linear regression  that incorporates the effect of the number of followers on popularity
\begin{eqnarray}
	\log R(t)= \alpha_t+ \beta^1_t \log R(T)+ \beta^2_t \log D(T) \nonumber \\
	+ \beta^3_t \log d_0+ \sigma_t \xi_t, \nonumber 
\end{eqnarray}
where $D(T)= \sum_{i: t_i<T} d_i$ is the cumulative number of followers up to time $T$, $d_0$ is the number of followers for the original poster, and the parameters $\alpha_t$ and $\beta^{1, 2, 3}_t$ are  obtained by minimizing the squared error 
\begin{eqnarray}
        && E_t(\alpha_t, \beta^1_t, \beta^2_t, \beta^3_t)= \sum_{k=1}^{n_{\rm tr} } 
        ( \log R_k(t) - \alpha_t
        \nonumber \\         
       && \quad - \beta^1_t \log R_k(T) - \beta^2_t \log D_k(T) - \beta^3_t \log d_{0, k}  )^2. \nonumber
\end{eqnarray}
The variance $\sigma_t$ is then determined by the maximum likelihood estimator $\hat{\sigma}_t^2= E_t(\hat{\alpha}_t, \hat{\beta}^1_t, \hat{\beta}^2_t, \hat{\beta}^3_t)/ n_{\rm tr}$, where $\hat{\alpha}_t$, $\hat{\beta}^{1, 2, 3}_t$,
 and $\hat{\sigma}_t^2$ are the fitted values of $\alpha_t$, $\beta^{1, 2, 3}_t$, and $\sigma_t^2$, respectively. 
The cumulative number of retweet $R(t)$ is predicted by the unbiased estimator
\[	
        \hat{R}(t)= R(T)^{ \hat{\beta}^1_t} D(T)^{ \hat{\beta}^2_t} d_0^{ \hat{\beta}^3_t}  \exp( \hat{\alpha}_t+ \hat{\sigma}^2_t/2). 
\]
\subsubsection{Reinforced Poisson process (RPP)~\cite{Shen2014,gao2015}. }
For the third model, we adapted a recent method, which is based on a time-dependent Poisson process, where the retweet rate $\lambda(t)$ is defined as
\[
	\lambda(t)= c f_{\gamma }(t) r_{\alpha}(R), 
\]
where $f_{\gamma }(t)= t^{-\gamma }$ describes the aging effect, $r_{\alpha}(R)= \epsilon+ \frac{1- e^{- \alpha (R+1)}}{1- e^{-\alpha}}$ is a reinforcement mechanism associated to the multiplicative nature of the spreading, and $R$ is the cumulative number of retweets at time $t$. The model parameters $\{ c, \gamma, \alpha \}$ are determined by maximizing the likelihood function~\cite{gao2015}. 
The log-likelihood function is maximized by the gradient descent method, and the iteration  terminated when a convergence criterion is satisfied, i.e., the relative change in the parameters is lower than $10^{-4}$. The learning rate for the gradient method is set to $10^{-5}$ and the parameters are optimized in the  range suggested in \cite{gao2015}, that is $1.5 \leq \gamma \leq 3.5$ and $0.001 \leq \alpha \leq 0.1$. 

After fitting the parameters, the cumulative number of retweets is evaluated from the expectation of the Poisson process, 
\[
	\frac{dR}{dt}= \lambda(t),
\]
which can be solved exactly
\[
        R(t)= ( \log(1+e^x)- x - \log \tilde{\epsilon}- \alpha )/ \alpha,
\]
with
\begin{eqnarray}
        x(t)=\frac{ \tilde{\epsilon} c \alpha (T^{1-\gamma}- t^{1-\gamma}) }{(1-\gamma) (1- e^{-\alpha})}- (R(T)+1) \alpha \nonumber \\
        - \log( \tilde{ \epsilon}- e^{- \alpha (R(T)+1)}), \nonumber
\end{eqnarray}
and $ \tilde{\epsilon}= 1+ \epsilon(1- e^{-\alpha})$. This expression is then used to predict the cumulative number of retweets. 
\subsubsection{SEISMIC~\cite{zhao2015seismic}.}
This fourth method has recently been proposed for predicting the final number of retweets
\cite{zhao2015seismic}
\[
        \hat{R}(\infty)=  R(T) + \alpha_T \frac{\hat{p}(T) \Delta D(T) }{1- \beta_T \hat{p}(T)},
\]
where $\hat{p}(T)$ is the infectious rate at the end of observation window $T$ and $\Delta D(T)= \sum_{i: t_i<T} d_i (1- \Phi(T-t_i) )$. 
The infectious rate $\hat{p}(T)$ is estimated by a kernel estimator, and their hyper-parameters are $\alpha_T= 0.326$, $\beta_T= 20$~\cite{zhao2015seismic}. 
Note that while the information diffusion model behind SEISMIC is a Hawkes process related to the one of TiDeH, its predictor is based on a Galton-Watson type branching process, whose parameters are fitted by the Hawkes process. In contrast, TiDeH also uses Hawkes process for the prediction of the future retweet activity. 
As it is designed, SEISMIC can only be applied  to predict the final number of retweets, not for the future time course of retweet activity. 
\subsection{Prediction results}
We now compare the prediction accuracy of the proposed method (TiDeH) with that of the three methods LR, LR-N and RPP. A comparison with SEISMIC is also performed when possible. 

\begin{figure}[h!]
  \centering
  \includegraphics[width=7cm]{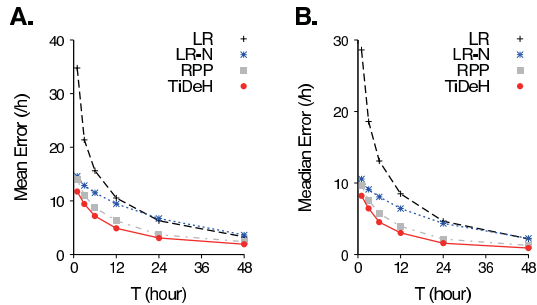}
  \caption{Comparison of prediction performance: Dependence of the error on observation time $T$. LR: Linear regression, LR-N: Linear regression with the number of followers, RPP: Reinforcement poisson process, TiDeH: proposed model. We predicted the retweet activity up to $T_{\rm max}= 168$ hours from the original post with the window size $\Delta_{\rm pred}= 4$ hours.}
  \label{fig:Comp_pred_T}
\end{figure}

\begin{figure}[h!]
  \centering
  \includegraphics[width=7cm]{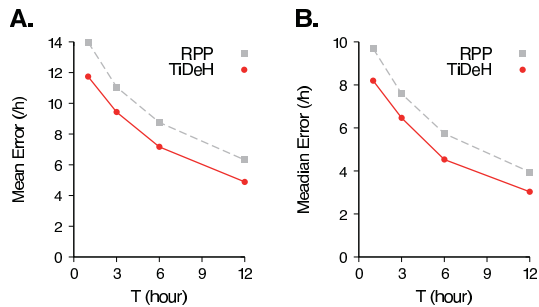}
  \caption{Magnified view of Fig.~\ref{fig:Comp_pred_T}. Prediction performance of the best two methods (RPP and TiDeH) were shown.}
  \label{fig:Comp_pred_TMag}
\end{figure}

%  Fig.9: Change $T$
First, we have examined the dependency of the prediction performance on the observation time $T$. To do so, we have performed a 5-fold cross validation test, except for RPP as it does not require training for the prediction. 
As shown in Figure~\ref{fig:Comp_pred_T}, TiDeH performs best in all the regimes, from short (1 hour) to long (48 hours) observation times, followed in order of accuracy by RPP, LR-N, and LR. 
In general, methods based on point processes (TiDeH and RPP) perform significantly better those based on linear regressions (LR-N and LR). 
%  Fig.10: Change $T$ for shot observations
We also observe that the errors increase when the observation time is decreased, and that this increase in  error is minimal for TiDeH. 
Figure~\ref{fig:Comp_pred_TMag} is a magnified view of Figure~\ref{fig:Comp_pred_T} clearly showing that TiDeH outperforms RPP, with a systematic improvement of accuracy of around  20 \%. On average, the error of TiDeH is 17.9 \% (mean error) and 21.7 \% (median error) smaller than that of RPP. Let us also note that LR-N performs much better than LR for  short observation times, confirming that network information, here the number of followers, is a key ingredient for prediction improvement. 
%  Change $DT

\begin{figure}[h!]
  \centering
    \includegraphics[width=7cm]{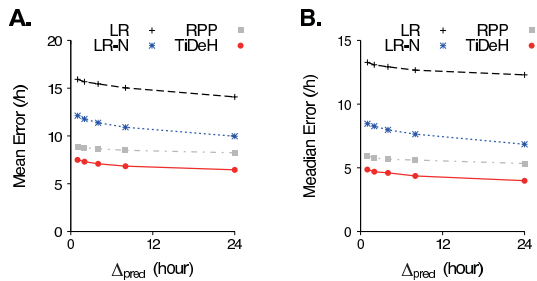}
  \caption{Comparison of prediction performance: Dependence of the error on time resolution $\Delta_{\rm pred}$. The abbreviations of the methods are the same as Fig.~\ref{fig:Comp_pred_T}. The observation time $T$ is fixed to 6 hours.}
  \label{fig:Comp_pred_DT}
\end{figure}

As a second step, let us consider the impact of the window size  $\Delta_{\rm pred}$ on prediction performance. Figure~\ref{fig:Comp_pred_DT} shows a similar pattern as above, with TiDeH the best predictor over all time scales, from precise (1 hour) to coarse (1 day) predictions, followed by RPP, LR-N, and LR. 
In general, the dependency of  the error on the window size is weak, and the error slightly decreases when the window size is increased, possibly because the observation time is not sufficient to learn the retweets dynamics with a greater accuracy and/or the retweet dynamics has characteristic times larger than 1 day. 
 
\begin{figure}[h!]
  \centering
    \includegraphics[width= 8cm]{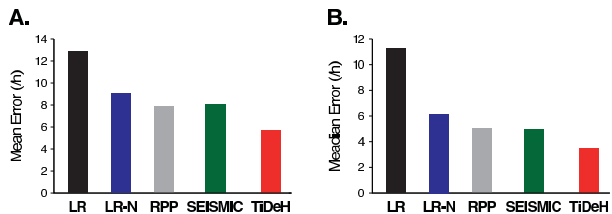}
  \caption{Comparison of prediction performance for the final number of retweets ($T= 6$ hours, $T_{\rm max}= 168$ hours). The abbreviations of the methods are the same as Fig.~\ref{fig:Comp_pred_T}. }
  \label{fig:Comp_Final}
\end{figure}

%  Final Number of Retweet
Finally, we estimate the prediction performance of TiDeH for a standard objective function, the final number of retweets. In addition to the three baseline method, we also compare its performance with the fourth baseline, SEISMIC. %  The hyper-parameters were adopted from their paper, i.e., $\alpha_T= 0.326$ and $\gamma_T n_*= 20$. 
Figure~\ref{fig:Comp_Final} shows that TiDeH provides again the most accurate predictions for the final number of retweets. 
 In terms of the mean and median error, we observe an improvement of around 30 \% over the two runners up (SEISMIC and RPP).

% \clearpage

\vspace{5cm}
%%
%%    
%%    Section  7
%%   	\input{sec7-conclusion.tex}
\section{Conclusion and Future work}
In this work, we have introduced TiDeH, a framework based on self-exciting point processes to predict the future time evolution of the popularity of a tweet. The method is based on the calibration of a model for information diffusion in social networks, which incorporates network information, circadian rhythms of online activity and aging of information. By doing so, the model provides a description based on absolute times, that is the time of the day, and relative time, that is the time since the previous triggering event, with a yet small number of parameters.
As compared to previous models, our approach also has the advantage of mathematical consistency, as the modeling and the prediction tasks are performed in the same framework, and leads to a systematic improvement of accuracy in a wide range of  time scales. Interestingly, our model also outperforms state-of-the-art methods to estimate the final number of retweets of a tweet, which emphasizes the importance of an appropriate modeling to solve prediction task.

Here, we have focused on the popular tweets that have more than 2,000 retweets, but the majority of information cascades on social networks are significantly shorter. It would be interesting to develop a parameter optimization technique for shorter data to overcome the limitation. 
Potential extensions of our work include a more detailed circadian activity by enriching the proposed model with higher harmonics and incorporating additional network information, such as correlations between number of followees and number of followers. 
% and the optimization of the model parameters by using machine learning techniques.
%  Other applications could take advantage of the temporal nature of the method in order to estimate in real time deviations of a stream of data from the predictions, and the study of hashtag dynamics, where endogeneous events continuously sustain online activity.
\vspace{0.5cm}

\section*{Acknowledgments}
This study was supported by JSPS KAKENHI Grant Number 25870915 to RK, and ARC and the Belgian Network DYSCO (Dynamical Systems, Control, and Optimismtion), funded by the Interuniversity Attraction Poles Programme to RL. This work is also the fruit of a Bilateral Joint Research Projects between JSPS, Japan, and F.R.S.-FNRS, Belgium. We thank Takaaki Aoki for stimulating discussions and anonymous reviewers for helpful comments. 

%  \bibliography{ref}% Produces the bibliography via BibTeX.

\end{document}